\documentclass[11pt,twocolumn,oneside]{article}
\usepackage[colorlinks,citecolor=blue,linkcolor=blue,urlcolor=blue]{hyperref}
\usepackage[a4paper,top=3cm,bottom=3cm,left=1.7cm,right=1.7cm]{geometry}
\usepackage[babel]{csquotes}
\usepackage{graphicx}
\usepackage{amsmath,amssymb}
\usepackage{url}
\usepackage{mathrsfs}
\usepackage{enumerate}
\usepackage{titlesec} 
\usepackage{authblk} 
\usepackage{abstract}
\usepackage[toc,page]{appendix}
\usepackage{cite}
\usepackage[margin=4pt, font=footnotesize, labelfont=bf, labelsep=period]{caption}	[2004/07/16]	
\usepackage{flushend}
\usepackage{balance}
\usepackage{cuted}
\usepackage{subfigure}
\usepackage[normalem]{ulem}

\def\be{\begin{equation}}
\def\ee{\end{equation}}

\newcommand{\dd}{ \mathrm{d}}

\renewcommand\thesection{\Roman{section}}

\titleformat{\section}{\large\scshape\bfseries\centering}{\thesection.}{.7em}{}
\titleformat{\subsection}{\scshape\bfseries}{\thesubsection.}{.7em}{}

\title{\Large{\textbf{\textsc{ On the volume inside old black holes}}}}

\author{Marios Christodoulou\thanks{christod.marios@gmail.com}
}
\author{Tommaso De Lorenzo\thanks{tommaso.de-lorenzo@cpt.univ-mrs.fr}
}

\affil{\small\textit{Aix Marseille Univ, Universit\'e de Toulon, CNRS, CPT, Marseille, France}}

\begin{document}
\twocolumn[
\vspace{-7ex}\maketitle \vspace{-5ex}%
\begin{list}{}{\leftmargin=3em\rightmargin=\leftmargin}\item\relax
\small\textbf{\textsc{Abstract.}} 
Black holes that have nearly evaporated are often thought of as small objects, due to their tiny exterior area. However, the horizon bounds large spacelike hypersurfaces. A compelling geometric perspective on the evolution of the interior geometry was recently shown to be provided by a generally covariant definition of the volume inside a black hole using maximal surfaces. In this article, we expand on previous results and show that finding the maximal surfaces in an arbitrary spherically symmetric spacetime is equivalent to a $1+1$ geodesic problem. We then study the effect of Hawking radiation on the volume by computing the volume of maximal surfaces inside the apparent horizon of an evaporating black hole as a function of time at infinity: while the area is shrinking, the volume of these surfaces grows monotonically with advanced time, up to when the horizon has reached Planckian dimensions. The physical relevance of these results for the information paradox and the remnant scenarios are discussed.

\end{list}\par\vspace{3ex}%
]
\saythanks

\bigskip    
  
\section*{Introduction} \label{sec:intro}

Since the mid-1970s, the information-loss paradox \cite{hawking:1976breakdown} has been at the center of a heated debate. The fate of the large amount of information fallen inside the hole is the main topic of several resolution proposals in the literature (for a --non-exhaustive-- review see \cite{hossenfelder2010conservative} and references therein).

In the setting in which the semi-classical approximation behind Hawking's computation remains valid up to the very late stages of the evaporation, and quantum gravitational effects play an important role only in the strong curvature regime by ``smoothing-out'' the singularity \cite{Ashtekar:2005cj}, a natural possible outcome is the formation of a \emph{remnant}: a final minuscule object that stores all the information needed to purify the external mixed state \cite{Aharonov:1987tp,Giddings:1993vj} (see \cite{Chen:2014jwq} for a recent review). 

The tiny mass and \emph{external} size of such objects are central to objections against both the existence of remnants (infinite pair production--see \cite{Giddings:1993km} and references therein--) and their impossibility of storing inside the large amount of information. The naive intuition of ``smallness", however, can be very misleading since a remnant contains spatial hypersurfaces of very large volume, see for instance \cite{AbhayILQGS,Perez:2014xca}.

\bigskip 
Once a horizon forms, surfaces of increasingly large volume start to develop. This characteristic is naturally captured by the manifestly coordinate independent definition of volume employing maximal surfaces  recently proposed by Carlo Rovelli together with one of the authors in \cite{Christodoulou:2014yia}, where it was applied to the interior of static black holes \footnote{Other definitions for the volume have been proposed elsewhere \cite{Parikh:2006aa,1742-6596-33-1-044,DiNunno:2008id,Ballik:2010aa,Cvetic:2011aa,Gibbons:2012aa,Ballik:2013aa}.}.

For asymptotically flat geometries, this volume can be parametrized with the advanced Eddington-Finkelstein time $v$ and is denoted as $V(v)$. In the interior of a static spherically symmetric black hole of mass $m_0$ formed by collapsing matter, the volume grows monotonically with $v$ and is given at late times $v \gg m_0$ by 
\begin{equation} \label{eq:schvolume}
V(v) \approx C\, m_0^2\, v
\end{equation} 
where $C=3 \sqrt{3}\,\pi$ for the uncharged case \footnote{The Reissner-Nordstr\"om spacetime, in which case $C$ depends on the charge $Q$, was studied in the Appendix of \cite{Christodoulou:2014yia} and similar results hold also for AdS black holes \cite{Ong:2015tua}. The Kerr case is considered in \cite{Bengtsson:2015zda}. }.

In this article, we expand upon the results in \cite{Christodoulou:2014yia} and show that the conclusions in that work extend to the case of an evaporating black hole. The volume of maximal surfaces bounded by the shrinking apparent horizon monotonically increases up to when its area has reached Planckian dimensions. Specifically, we show that, at any time, there exists a spacelike maximal surface with proper volume approximately given by \eqref{eq:schvolume} (where $m_0$ is now the initial mass), that connects the sphere of the apparent horizon at that time to the center of the collapsing object before the formation of the singularity \footnote{An argument for the persistence of the large volume in the evaporating case was discussed in \cite{Ong:2015dja}.}. The final remnant hides inside its external Planckian area a volume of order $(m_0/m_P)^5 \, l_P{}^3$.

\bigskip

We first review and clarify some aspects of the discussion given in \cite{Christodoulou:2014yia} and generalize the results presented there so that they may be used in an arbitrary spherically symmetric spacetime. In Section \ref{sec:maxSurAs2Dgeod} and the Appendix, we prove the technical result that finding the spherically symmetric maximal surfaces is equivalent to solving a two dimensional geodesic problem. In Section \ref{sec:definition} we review the definition of volume and discuss the analogy between the Minkowski and the Schwarzschild case in order to illustrate its geometric meaning. In Section \ref{sec:remnantvolume} we examine the evaporating case and calculate the volume enclosed in the horizon as a function of time at infinity. We close with a discussion on the physical relevance of our result with respect to the debate on the fate of information in evaporating black holes.

\section{Maximal surfaces as a $1+1$ geodesic problem} \label{sec:maxSurAs2Dgeod}

A general spherically symmetric spacetime can be described by a line element
\be
\dd s^2 = g_{\alpha \beta}\dd x^\alpha \dd x^\beta = g_{A B}\dd x^A \dd x^B + r^2 \dd \Omega^2 
\ee 
with $\dd \Omega^2 = \sin^2\!\theta\, \dd \phi^2 + \dd \theta^2 $. We use the notation $x^\alpha = \{x^0,r,\theta,\phi\}$ and $x^A = \{x^0,r\}$.

Spherically symmetric hypersurfaces $\Sigma$ can be parametrically defined via a coordinate $\lambda$: 
\be
\dd s^2_\Sigma =  (g_{A B} \dot{x}^A \dot{x}^B) \,\dd \lambda^2 + r^2 \dd \Omega^2 
\ee 
where $x^A = x^A(\lambda)$ and $\dot{x}^A \equiv \frac{\dd}{\dd \lambda}  x^A \!(\lambda)$. We have $\Sigma \sim \gamma \times S^2$, with $\gamma : \lambda \rightarrow x^A(\lambda)$ being a curve in the $x^0\textit{-}\, r$ plane. We denote as $y^a=\{\lambda,\phi,\theta\}$ and $h_{ab} = e^\alpha_a e^\beta_b g_{\alpha \beta}$ the coordinates and the induced metric on $\Sigma$ respectively, where $e^\alpha_a = \frac{\partial x^\alpha}{\partial y^a}$ provides a basis of tangent vectors on $\Sigma$.

We look for the stationary points of the volume functional: 
\begin{eqnarray}
V[\Sigma] &=& \int_\Sigma\!\dd y^3 \sqrt{\det h_{ab}} \nonumber \\ 
&=& 4 \pi \int_{\gamma}\dd \! \lambda \left(r^4 g_{A B} \dot{x}^A \dot{x}^B \right)^{1/2}  \nonumber \\
&=& 4 \pi \int_{\gamma} \dd \! \lambda \left(\tilde{g}_{A B}  \dot{x}^A \dot{x}^B\right)^{1/2} \;,
\end{eqnarray}
where $\Sigma$ are spherically symmetric surfaces bounded by a given sphere $\partial \Sigma$.

Thus, the extremization of $V[\Sigma]$ is equivalent to the 2D geodesic problem for the auxiliary metric $\tilde{g}_{A B}= r^4 \, g_{A B}$. That is, $\gamma$ is a solution of 
\be \label{eq:geodEq}
\dot{x}^A \tilde{\nabla}_A \dot{x}^B = e^A_\lambda \tilde{\nabla}_A e^B_\lambda = 0    
\ee 
where $\tilde{\nabla}$ is the covariant derivative in $\tilde{g}_{A B}$ and $\lambda$ has been chosen to be an affine parameter on $\gamma$ with respect to $\tilde{g}_{A B}$. 

\bigskip

The stationary points of $V[\Sigma]$ solve the ``Plateau's problem'' or ``isoperimetric problem'' for $\partial \Sigma$. In a Euclidean context these are local minima, while in the Lorentzian context they are local maxima. It is simple to show that if the trace $K=K_{\alpha \beta} g^{\alpha \beta}$ of the extrinsic curvature of a hypersurface vanishes, the variation of the volume functional is automatically zero (see for instance \cite{Baumgart}). For this reason, in the Lorentzian context, surfaces with $K=0$ are called maximal surfaces. 

It is the authors understanding that a general proof of the opposite statement, namely that for arbitrary spacetimes extremizing $V[\Sigma]$ for a given $\partial \Sigma$ yields $K=0$ surfaces, is missing. Several precise proofs exist in the mathematical relativity literature (see for instance the seminal papers \cite{Choquet1979,Marsden1980}), that typically rely on energy conditions or other restrictions on the metric or on the surfaces. ``Physicist'' demonstrations can be found in the $3+1$ literature \cite{Gourgoulhon,Baumgart}. 

For completeness, we prove in the Appendix that, for an arbitrary metric $g_{AB}$, any surface $\Sigma \sim \gamma \times S^2$, with $\gamma$ being a solution of \eqref{eq:geodEq}, has $K=0$. From well known theorems about the geodesic equation, this also guarantees the local existence of maximal surfaces, see also \cite{CorderoCarrion:2011rf}. 

\vspace{1ex}
\centerline{\rule{2cm}{1.5pt}}
\vspace{1.5ex}

The physical relevance of maximal surfaces has long been recognised in diverse disciplines ranging from problems in mathematical physics \cite{Rassias} to architecture and the beautiful tensile structures of Frei Otto \cite{FreiOtto}. In general relativity, their usefulness for numerically solving Einstein's equations is reflected in the popular ``maximal slicing'' \footnote{The family of surfaces discussed in the next section includes the surfaces used for maximal slicing, but keep in mind that we do not restrict ourselves to surfaces satisfying the ``singularity avoidance'' or the ``nowhere-null'' condition. In fact, half of each family of $K=0$ surfaces we will study end at the singularity and become null there.} (see for instance \cite{Gourgoulhon} and references therein), which in a sense generalizes the slicing of a Newtonian spacetime by constant (absolute) time surfaces. 

Common notions of volume implicitly use maximal surfaces. These include the everyday meaning of volume, the special relativistic proper volume and the volume of the Universe, where the latter habitually refers to the proper volume of the $t=const.$ surfaces of the Friedmann-Robertson-Walker metric: spherically symmetric maximal surfaces. 

\section{Review of the volume definition} \label{sec:definition}
\begin{figure*}[ht]
\centering
\subfigure{
\centering
    \includegraphics[width=0.41\textwidth, keepaspectratio]{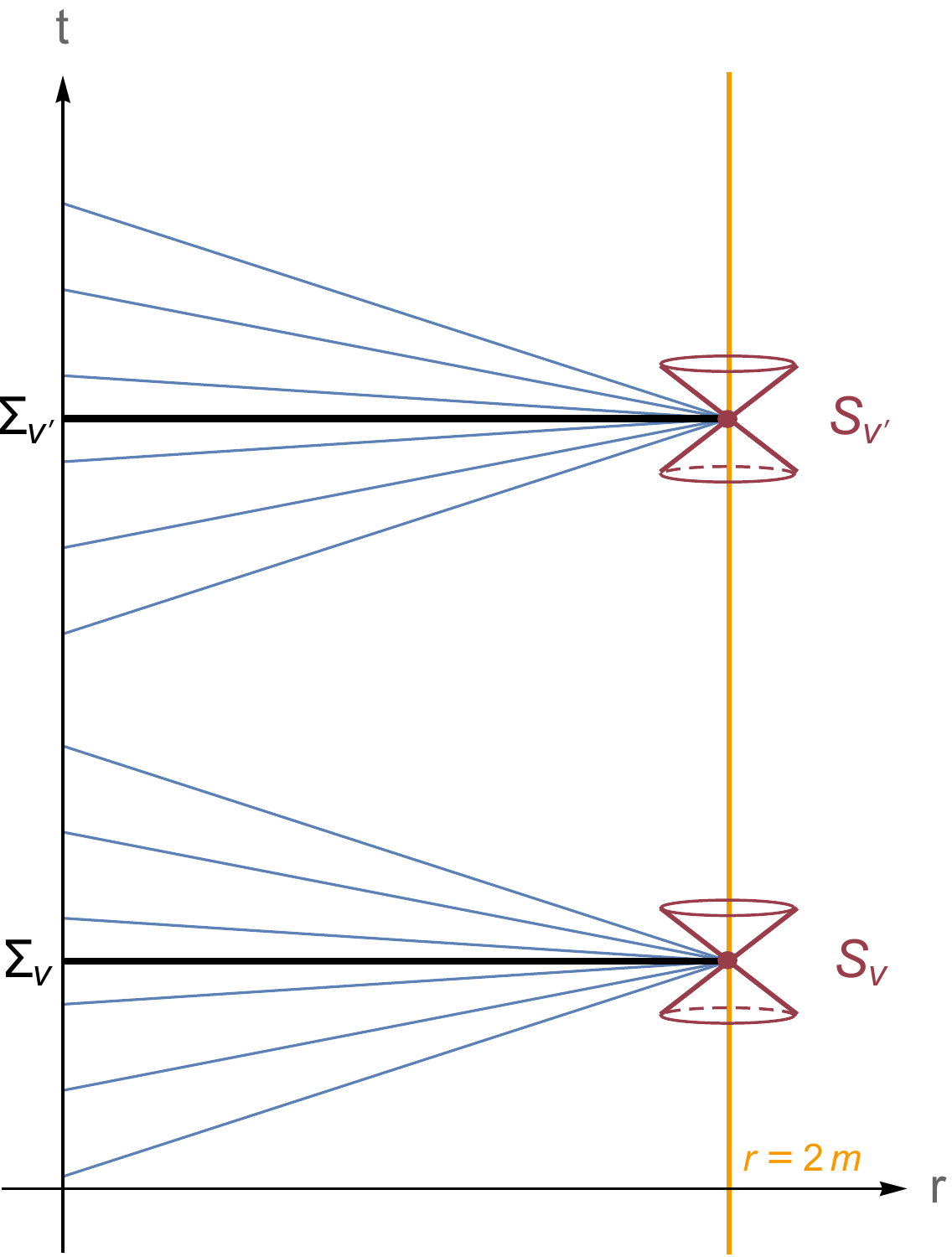}
    \label{fig:Minkowski}
}\qquad\qquad
\subfigure{
\centering
    \includegraphics[width=0.41\textwidth, keepaspectratio]{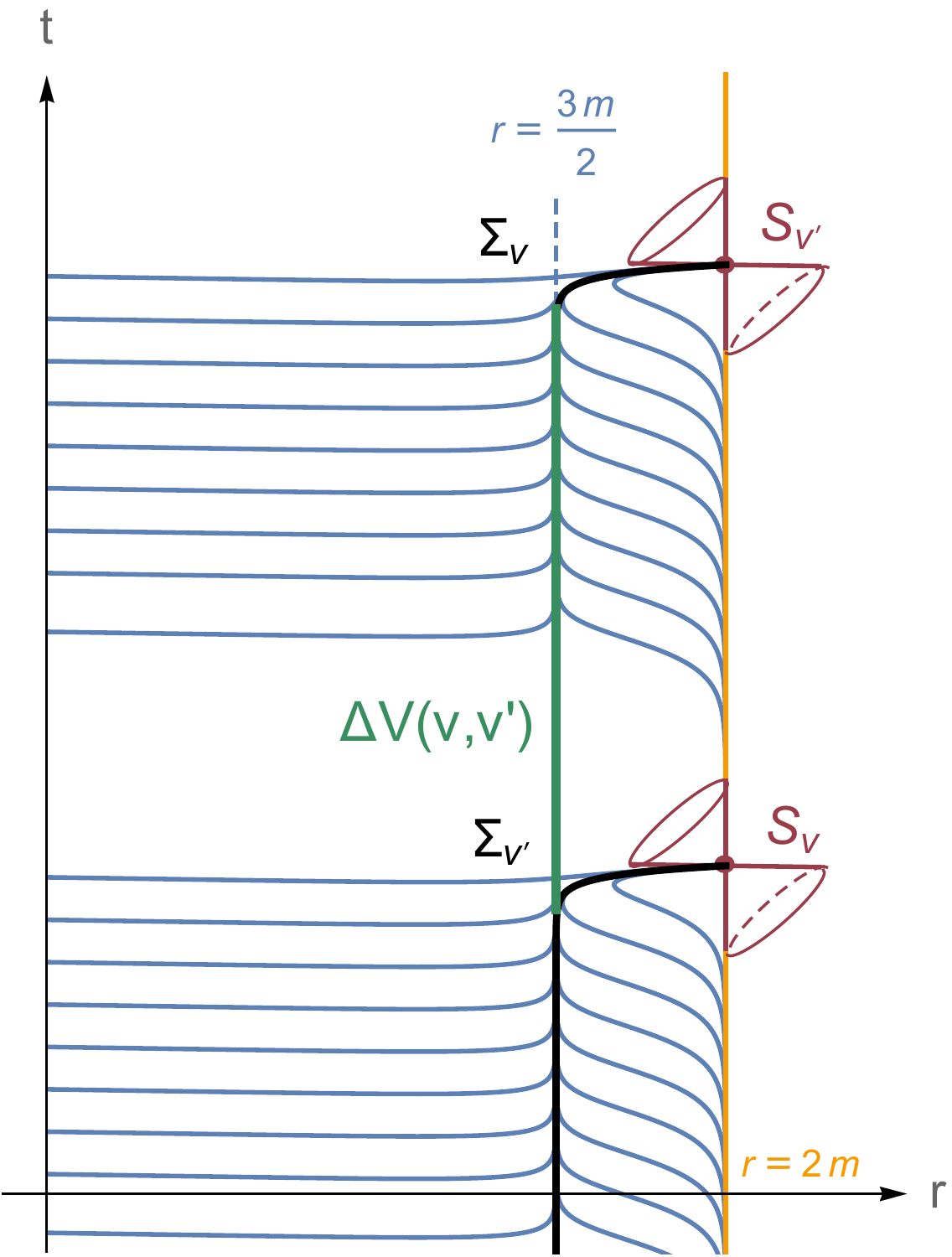}
\label{fig:Schwar}
}
\caption{{\bf Left:} Maximal surfaces (blue lines) inside a two-sphere in flat Minkowski spacetime. The largest is the $t={\rm const.}$ (bold black lines) defining its inertial frame. {\bf Right:} Maximal surfaces (blue lines) inside a two-sphere on the horizon of a static black hole. Apart from the transient part connecting it to the horizon, the largest surface (bold black lines) lies on the limiting surface $r=3/2m$. The volume \emph{difference} between the spheres $S_v$ and $S_{v'}$ is finite and given by \eqref{eq:SchVac}.}
\label{fig:mink_VS_schw}
\end{figure*}
    The volume definition given in \cite{Christodoulou:2014yia} can be stated as follows: \emph{the volume inside a sphere $S$ is defined as the proper volume of the maximal spherically symmetric surface $\Sigma$ bounded by $S$, which has the largest volume amongst all such $\Sigma$.} Note that this is a geometric statement and as such it is manifestly generally covariant.

\begin{figure}[t]
\centering
\includegraphics[width=0.35\textwidth]{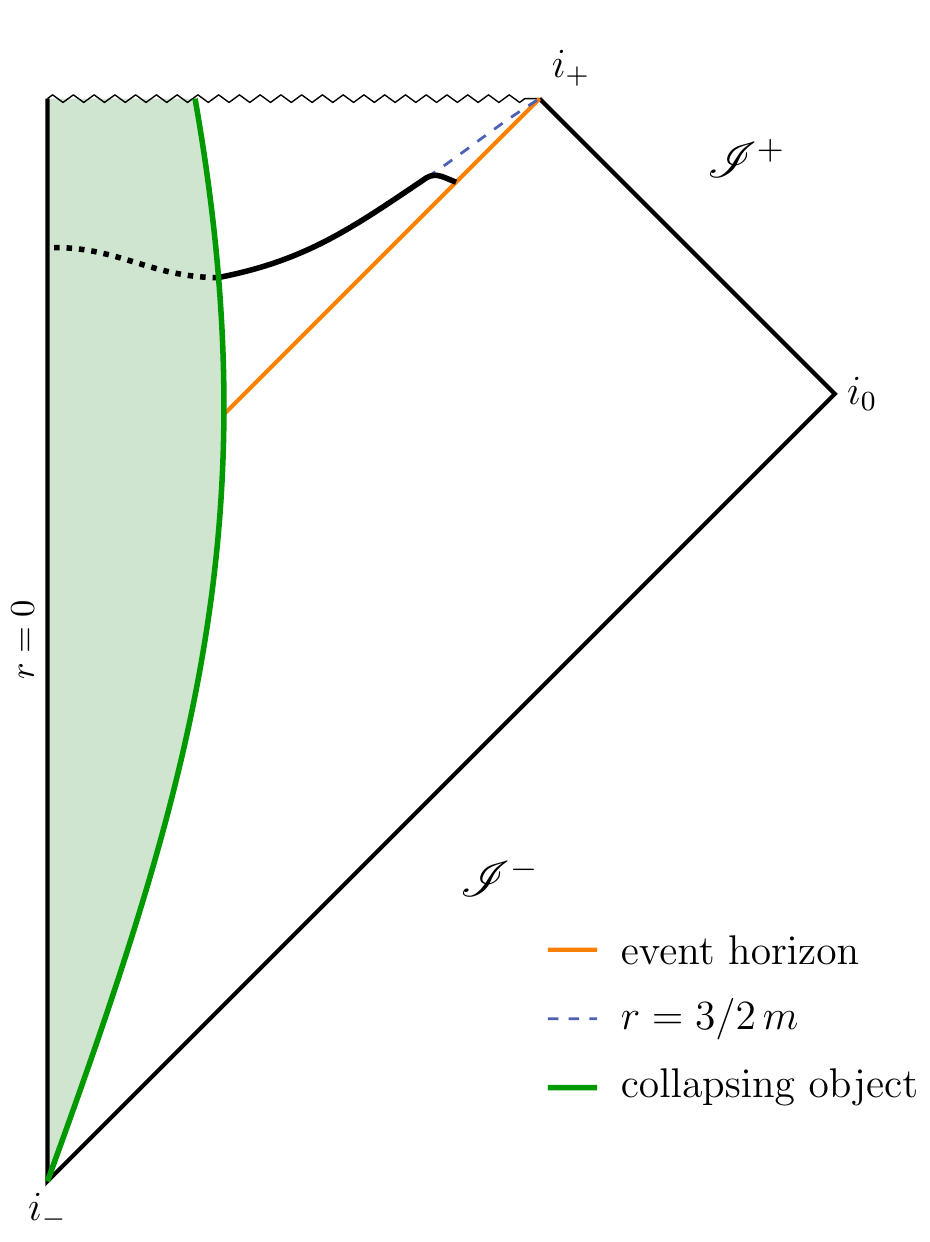}
\caption{Penrose diagram illustrating the surface defining the volume (black curve) in the case of a black hole formed by collapse. The details of the surface in the interior of the collapsing object (dotted curve) will depend on the specific metric use to describe the latter. For Oppenheimer-Snyder and null massive shell collapses, this contribution to the volume is of the order $m^3$.}
\label{fig:collapse}
\end{figure}

 In order to illustrate its geometric meaning, we examine in the rest of this section the analogy between the maximal surfaces of Minkowski spacetime and those of the Schwarzschild solution. The discussion is summarized in Figures \ref{fig:mink_VS_schw} and \ref{fig:collapse}.

 Using the advanced time $v=t+ \int \frac{\dd r}{f(r)}$, the geometry of the two spacetimes is described by 
\be \label{eq:vacLineEl}
\dd s^2= -f(r) \dd v^2+  2 \dd v \dd r +r^2 \dd \Omega ^2\;,
\ee 
with $f(r)=1$ and $f(r)=1- 2m/r$ respectively. Consider the sphere $S_v$ defined as the intersection of $r=2m$ and the ingoing radial null ray of constant $v$. It bounds a family of maximal surfaces, the solutions of \eqref{eq:geodEq} for different initial speeds. 

In Minkowski, these are the simultaneity surfaces of inertial observers, which are straight lines in the $t\textit{-}r$ plane. The one with the biggest volume, $\Sigma_v$, is that which defines the inertial frame of $S_v$. Its proper volume is what we call \emph{the} proper volume in special relativity; that is, $V_{\Sigma_v}=\frac43 \pi (2m)^3$. 

In Schwarzschild geometry, the maximal surfaces starting from $S_v$ approach the surface $r=3/2m$ (because of this behavior, $r=3/2m$ will be called ``limiting surface''), and become null either when they reach the singularity or when they asymptotically approach the horizon, except one that asymptotically becomes $r=3m/2$ \footnote{The existence of the limiting surface $r=3/2m$ was first pointed out in \cite{Estabrook:1973ue}. It is crucial for the singularity avoidance property of the maximal slicing, which is in fact comprised by the $\Sigma_v$ extended to infinity. Similar elongated surfaces are studied in numerical relativity \cite{Hannam:2006xw,Baumgarte:2007ht} and have been dubbed ``trumpet geometries'' \cite{Dennison:2014eta}. }. The proper volume of this surface is infinite.

This is a characteristic difference between the two geometries which underlines the common understanding that ``space and time exchange roles inside the hole''. Inside the sphere containing flat space, there are radial timelike curves of infinite length, while all radial spacelike curves have proper length at most equal to the radius of the sphere. Inside a black hole this is reversed: there are radial spacelike curves of infinite length, while radial timelike curves have proper time at most equal to $\pi m$. 

In the physical case of non-eternal black holes formed by collapse, the surface $\Sigma_v$ does not have infinite volume since it does not extend infinitely along $r=3/2m$. In fact, it connects the sphere at the horizon $S_v$ with the center of the collapsing object before the formation of the singularity, see Fig.~\ref{fig:collapse}. The surface in its interior will be given by solving \eqref{eq:geodEq} for the interior metric. For a collapse modeled by a null massive shell or \emph{\`a la} Oppenheimer-Snyder \cite{PhysRev.56.455}, the contribution to $V(v)$ will be of the order of that of the flat sphere $\sim m^3$. At late times $v >> m$, this contribution is negligible with respect to the one given by the main part lying on $r=3/2m$, and the volume is given by $\eqref{eq:schvolume}$.

This characteristic monotonic behaviour is perhaps best understood by extending the definition to the case of an eternal black hole. In this case we consider the volume \emph{difference} $\Delta V(v,v')$ between two spheres $S_v$ and $S_{v'}$ labeled by different times at infinity, in analogy to considering the proper time between any two points on a timelike curve that otherwise extends to arbitrary values of its affine parameter. 
   
In Minkowski, this difference is zero: the proper volume of the sphere of fixed radius remains constant. In Schwarzschild, by the translation invariance inside the horizon, $\Delta V(v,v')$ is given by the volume of the part of $\Sigma_{v'}$ that lies on the limiting surface $r=3m/2$ and does not overlap with $\Sigma_v$. \emph{Thus, this difference is finite, monotonically increasing and given by} 
\be \label{eq:SchVac}
\Delta V(v,v') = 3 \sqrt{3}\, \pi \, m^2 \, (v'-v)\;.
\ee
Notice that the result for a black hole formed by collapse, eq.~\eqref{eq:schvolume}, is nothing but the approximate version of the above equation with $v=0$.

The analysis presented in this section can be nicely extended to the case of an evaporating black hole to which we now turn our attention. 

\section{The volume of an evaporating black hole}\label{sec:remnantvolume}
The spacetime of an evaporating spherically symmetric black hole can be described by the Vaidya metric \cite{Vaidya:1951zz}, given by replacing $f(r)$ in \eqref{eq:vacLineEl} with $f(r,v)=1 - 2m(v)/r$. For our purposes it is sufficient to model the formation of the hole by the collapse of an ingoing null shell at the retarded time $v=0$, and the loss of mass due to evaporation by integrating the thermal power emission law \cite{Hawking:1974}. The resulting mass function is
\be\label{eq:mv}
m(v) = \Theta(v) \big(m_0^3-3B\, v\big)^{1/3} \;,
\ee
where $\Theta(v)$ is the step function, $B \sim 10^{-3}$ a parameter that corrects for back reaction \cite{PhysRevD.52.5857} and $m_0$ the mass of the shell. The spacetime has a shrinking timelike apparent horizon given by $r_{H}(v) = 2 m(v)$.

\begin{figure}[t]
\centering
\includegraphics[width=0.35\textwidth]{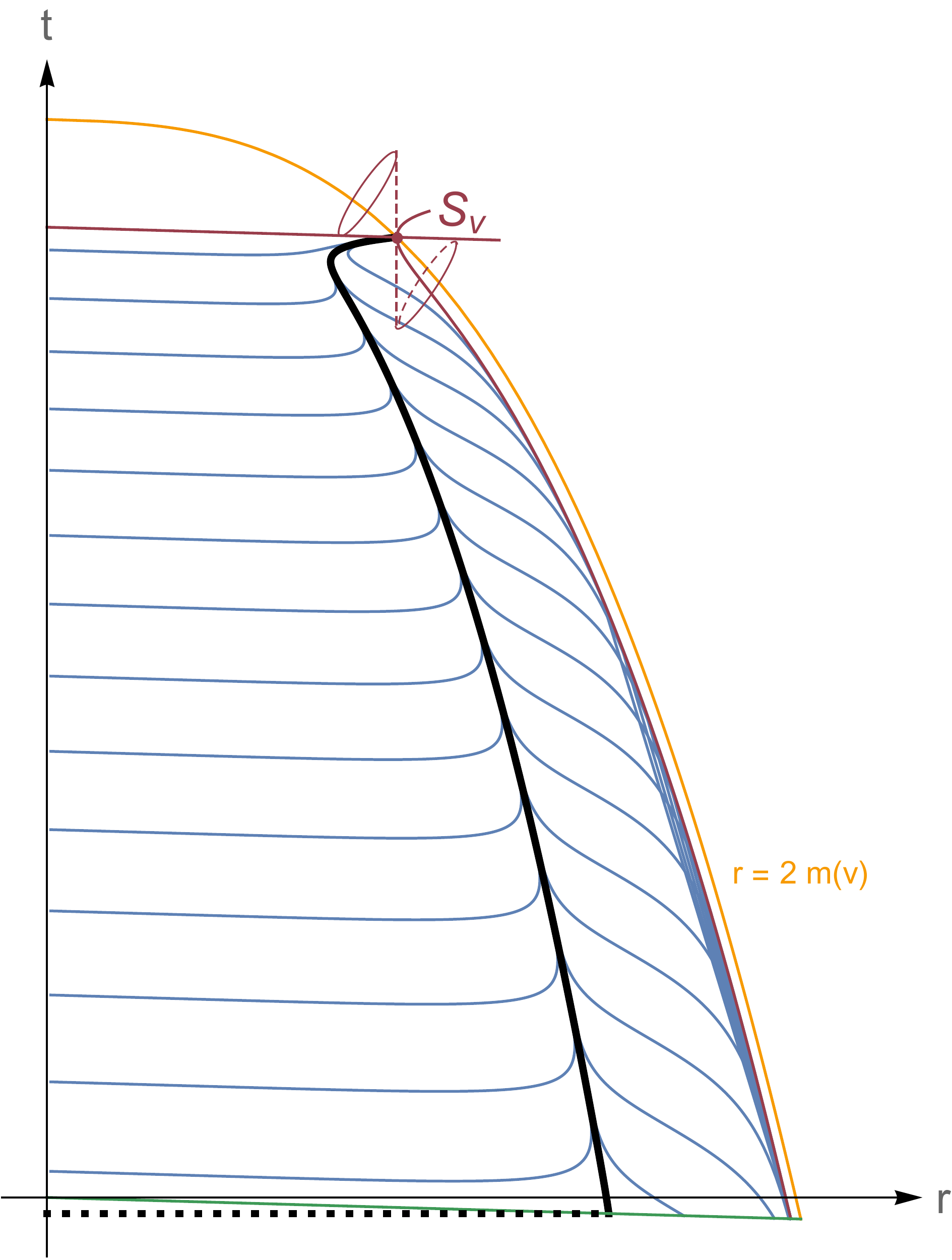}
\caption{Eddington-Finkelstein diagram of the two families of extremal volume surfaces (blue lines) inside an evaporating black hole formed by a collapsing object. The surface defining the volume is in bold black. Note the close analogy with the static case, compare with Fig.~\ref{fig:mink_VS_schw}.}
\label{fig:evaporating}
\end{figure}

By numerically solving \eqref{eq:geodEq}, we can draw the family of maximal surfaces for the spheres at the apparent horizon for different $v$. The situation, depicted in Figure \ref{fig:evaporating}, is in direct analogy with the non-evaporating case. There is again a limiting surface, persisting up to very late stages of the evaporation. Thus, as in the static case, the volume of the biggest maximal surface $\Sigma_v$ inside $S_v$ is the one connecting the latter to the center of the collapsing shell.

\begin{figure}[t]
\centering
\includegraphics[width=0.35\textwidth]{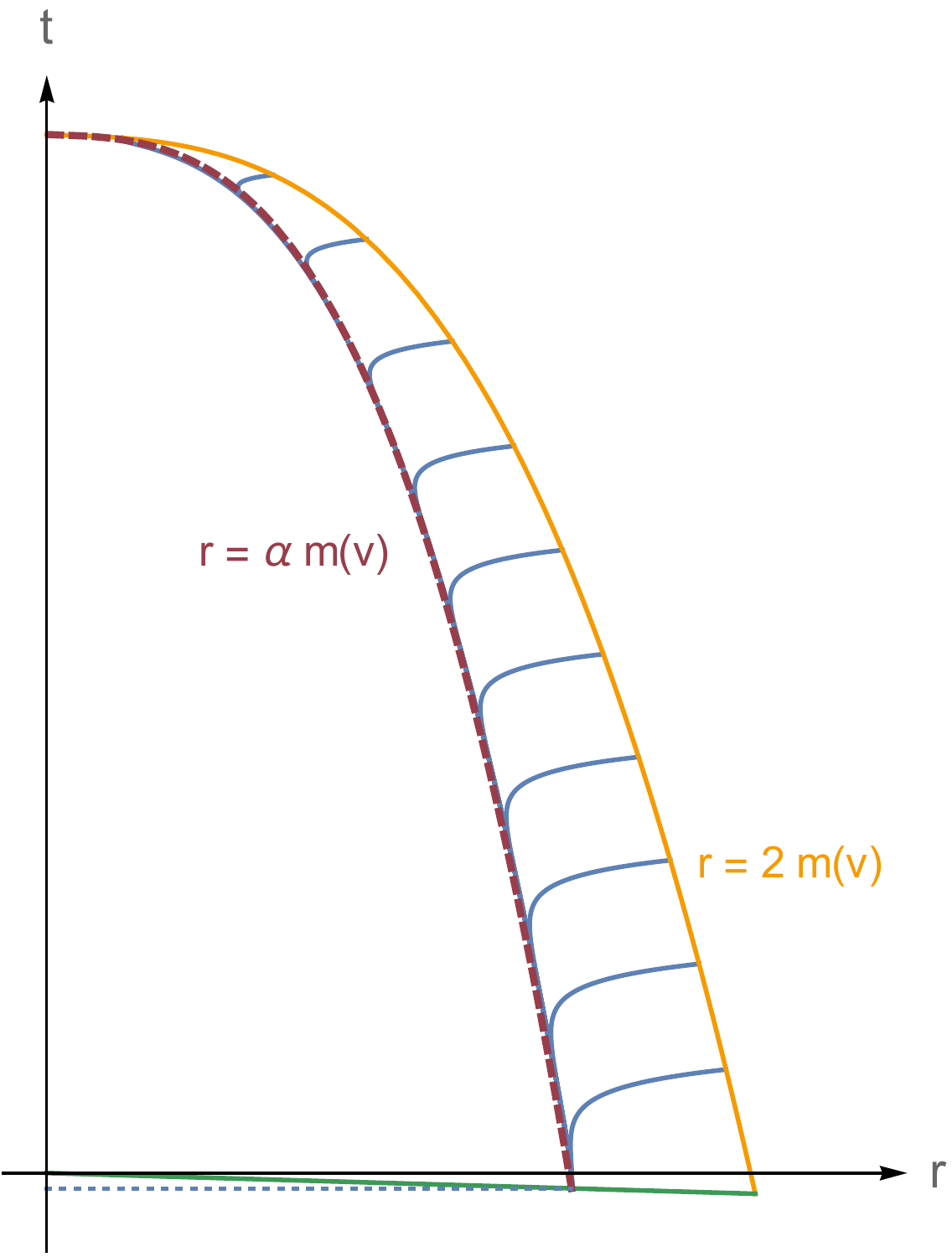}
\caption{The surfaces defining the volume enclosed in spheres at the apparent horizon of an evaporating black hole at different times (blue lines). The limiting surface lies close to $r=\alpha\,m(v)$, with $\alpha$ given by \eqref{eq:alpha} (dashed line). Here $m=10$ in Planck units.}
\label{fig:complete}
\end{figure}

We may get an estimate for the volume as a function of time and the initial mass as follows: we compute the volume of a surface $r=\alpha\, m(v)$ and find the $\alpha$ for which this is maximized:
\be\label{eq:alpha}
\alpha = \frac32 - \frac{45\, B}{8\, m_0^2} + O\left(\frac{1}{m_0^4}\right)\;.
\ee
Indeed, the limiting surface is very well approximated by $r=\alpha\, m(v)$ even for low masses, see Fig.~\ref{fig:complete}. 

Expanding the volume of $r=\alpha\, m(v)$ to leading order in $1/m_0$ we get: 
\be\label{eq:volapprox}
V(v) \approx 3\sqrt{3}\,\pi \,m_0 ^2 \,v \,\left( 1- \frac{9 \,B}{2\,m_0^2}\right) \;.
\ee

Thus, for large masses, we have again recovered \eqref{eq:schvolume}. 

A direct calculation shows that the surface $r(v) = \alpha\,m(v)$ ceases to be spacelike when the mass function takes the value
\begin{equation}
m \approx \left(3 \sqrt{B} - \frac{225\,B^{3/2}}{8\,m_0^2}\right) \,m_P < m_P / 10 \;.
\end{equation}
This provides an estimate for the regime of validity of eq.~\eqref{eq:volapprox}. Interestingly, the non-existence of large spacelike maximal surfaces appears to coincide with the regime in which the mass has become Planckian. 
These estimates agree with the numerical investigation of the actual surfaces, see Fig.~\ref{fig:complete}. We conclude that the volume increases monotonically, following the approximate behavior given in \eqref{eq:volapprox}, up to when its external area becomes Planckian. At this very late time, the internal volume is of order $m_0^5$ in Planck units.

\vspace{1ex}
\centerline{\rule{2cm}{1.5pt}}
\vspace{1.5ex}

Intuitively, the picture is the following: from the perspective of the maximal surfaces, collapse and horizon \textit{at any subsequent exterior time} are simultaneous, see Fig.~\ref{fig:complete}.
The exterior elapsed time corresponds inside the hole to the stretching of space, as given by \eqref{eq:schvolume}.
 
\subsection*{A few numbers}
Before closing this section, let us put the above in perspective: when a solar mass ($10^{30} \, kg$) black hole becomes Planckian (it needs $10^{55}$ times the actual age of the Universe), it will contain volumes equivalent to $10^5$ times our observable Universe, hidden behind a Planckian area ($10^{-70} \,m^2$). 

Perhaps more pertinent is to consider small primordial black holes with mass less than $10^{12}\, kg$. Their initial horizon radius and volume are of the order of the proton charge radius ($10^{-15}m$) and volume ($10^{-45}m^3$) respectively. They would be in the final stages of evaporation now, hiding volumes of about one liter ($10^{-3}m^3$).

\section{Remnants and the information paradox }\label{sec:discussion}

As was briefly discussed in the introduction, the results presented above can be relevant in the discussion about the loss-of-information paradox, particularly in the context of scenarios that assume the semiclassical analysis of quantum field theory on curved spacetimes to be valid in regions of low curvature and until near-complete radiation of the initial mass \footnote{Another potential application of this result is in black hole thermodynamics in view of recent results on the Von Neumann entropy associated to volumes \cite{Astuti:2016dmk}.}. Such scenarios disregard the possibility of having information being carried out of the hole by the late Hawking photons \cite{PhysRevLett.71.3743,Braunstein:2009my}, avoiding the recent firewall and complementarity debate \cite{Almheiri2013}. Another alternative that has recently aroused interest and is not considered here, is that a black hole may end its lifetime much earlier than near-complete evaporation by tunneling to a white-hole geometry. This is possible thanks to quantum gravitational effects that, due to the long times involved, can become important in low curvature regions outside the horizon \cite{Haggard:2014rza,DeLorenzo:2015gtx,TunnelTime} \footnote{An alternative scenario in which this process happens must faster by assuming faster-than-light propagation of a shock-wave from the bounce region is considered in \cite{Barcelo:2014npa,Barcelo:2015noa}.}. 

Consider then the setting in which the semi-classical approximation behind Hawking's computation remains valid up to the very end of the evaporation. The hole will completely evaporate and the information will unavoidably be lost, as originally suggested by Hawking \cite{hawking:1976breakdown}. While it seems intuitively reasonable for what appears to be a tiny object to decay away and disappear, it is compelling to ask what became of the macroscopic region inside.

Conversely, consider the additional hypothesis that quantum gravitational effects play an important role in the strong curvature regime by ``smoothing-out'' the singularity \cite{Ashtekar:2005cj}. When the mass becomes Planckian, the semi-classical approximation underlying Hawking's computation fails and the evaporation stops (see for instance \cite{Adler:2001vs}). The hole does not completely disappear and one can consider the possibility of having a minuscule object that stores all the information needed to purify the external mixed state: a remnant \cite{Aharonov:1987tp,Giddings:1993vj,Chen:2014jwq}. 

Standard objections against the remnant scenario such as the infinite pair production \cite{Giddings:1993km} and their impossibility in storing inside a large amount of information, rely on considering the remnant as a small object. Our result shows that the remnant is instead better understood as the small throat of an immense internal region, with a volume of the order of $m_0^5$. General Relativity naturally gives a ``bag of gold'' type description of the interior of a remnant, without the need of ad-hoc spacetimes that involve some ``gluing'' of geometries \cite{wheeler1964relativity,Hsu:2007dr}. Notice that the result of the previous section is insensitive to the details of the \emph{would-be-singularity} region since the limiting surface is in a relatively low-curvature region. 

In \cite{hossenfelder2010conservative,AbhayILQGS,Perez:2014xca} the authors suggest that a large available internal space could store a sufficient amount of very long wavelength modes that carry all the information needed to purify the external mixed state, albeit the available energy being of the order of a few Planck masses. The surfaces studied here are good candidates on which this idea could be tested \footnote{In \cite{Zhang:2015gda} it is argued that these surfaces do not store enough information for purification. However, in that work the Hawking temperature is assumed constant. The computed information is therefore the one stored in a static black hole, and it is not pertinent to this discussion.}. The details of the mechanism by which information would be stored have not, to our knowledge, been made precise; demonstrating this possibility is beyond the aim of this work and, in what follows, we assume this to be possible. 
 
We can identify two characteristically distinct possibilities for the evolution of the large interior region. The bulk of these large surfaces is causally disconnected from their bounding sphere on the horizon \cite{Bengtsson:2015zda}. They can remain causally disconnected from the rest of the spacetime, which may lead to a baby universe scenario \cite{Frolov:1988vj,Frolov:1989pf}. 

\begin{figure}[t]
\centering
\includegraphics[width=0.29\textwidth]{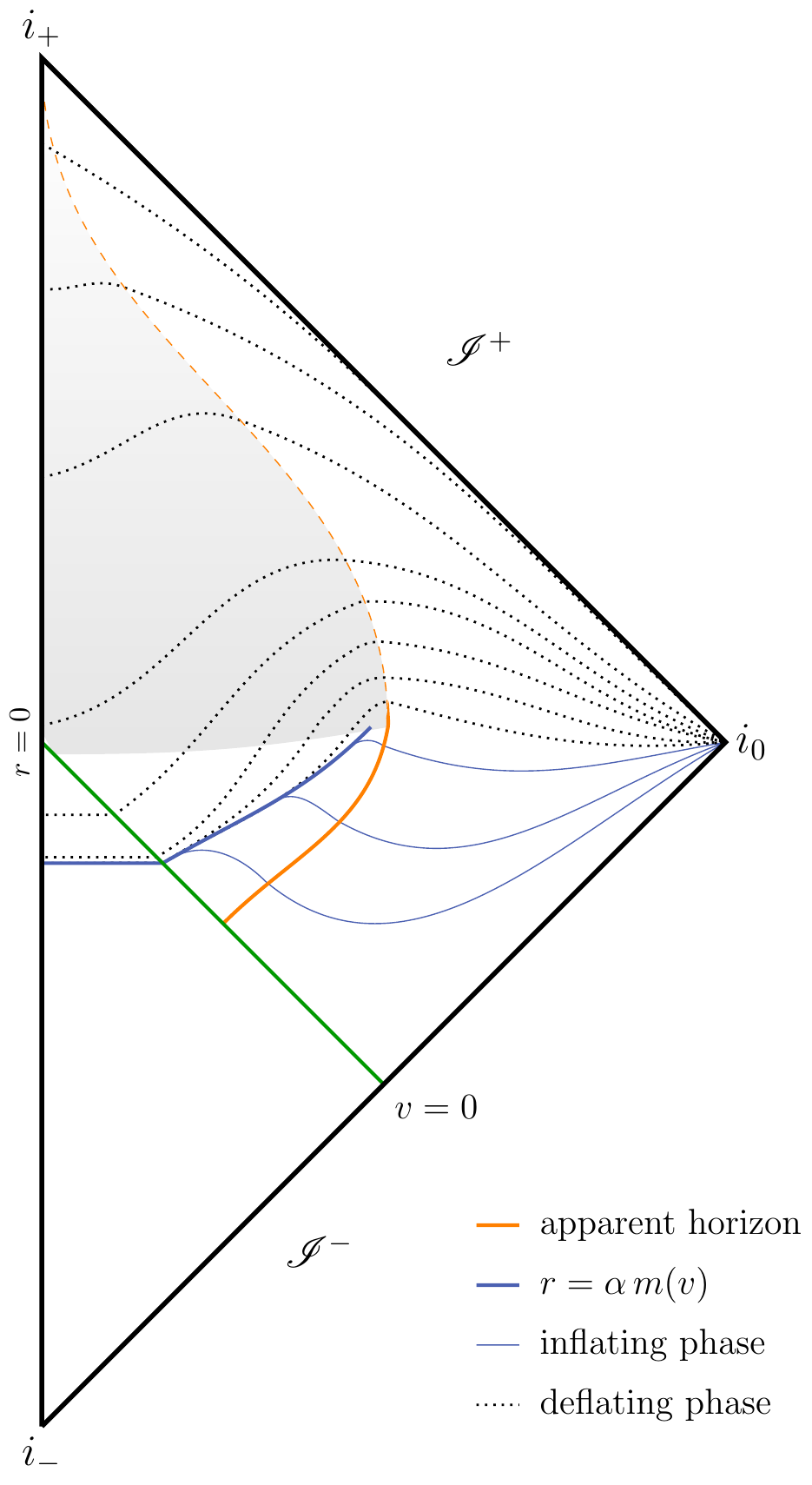}
\caption{Speculative evolution of maximal surfaces in the case of a long-lived remnant scenario. The volume acquired during the evaporation process (continuous surfaces) deflates back to flat space (dotted surfaces). This is expected to happen in a time of order $m_0^4$, during which all the information stored can be released.}
\label{fig:conjecture}
\end{figure}

On the contrary, quantum gravitational effects can modify the (effective) metric and bring these regions back to causal contact with the exterior, while deflating their volume, allowing for the emission of the purifying information to infinity (the information could also be coded in correlations with the fundamental pre-geometric structures of quantum gravity, as proposed in \cite{Perez:2014xca}). This scenario, where the inflating phase is followed by a slow deflating phase of the remnant, is sketched in Fig.~\ref{fig:conjecture}.

We expect this deflating process to be slow, in accordance with bounds on the purification time \cite{Carlitz:1986ng,Bianchi:2014bma} and the lifetime of long-lived emitting remnants, estimated to be of order $m_0^4$. The latter scenario can be made precise by constructing an effective metric describing this process through the evolution of maximal surfaces in the sense of Fig.~\ref{fig:conjecture}. It then suffices to numerically solve equation \eqref{eq:geodEq} in order to study the evolution.

\section*{Acknowledgments}
The authors thank Carlo Rovelli, Alejandro Perez and Thibaut Josset for the many discussions on this subject, Abhay Ashtekar for private communications that clarified aspects of the ideas in \cite{AbhayILQGS} and Thomas Baumgart for exchanges on maximal surfaces and their use in numerical relativity.

 We would like to thank Marina Konstantatou for pointing us towards applications of maximal surfaces in engineering and to the inspiring work of Frei Otto.

\medskip

M.C. acknowledges support from the Educational Grants Scheme of the A.G.Leventis Foundation for the academic years 2013-14, 2014-15 and 2015-16 as well as from the Samy Maroun Center for Space, Time and the Quantum.

\section*{Appendix}\label{app:Knull}
In the notation of Section \ref{sec:maxSurAs2Dgeod}, the mean extrinsic curvature is defined by
\be
K =\nabla_\alpha n^\alpha= h^{ab} e^\alpha_a e^\beta_b \nabla_\alpha n_\beta 
\ee 
where $\nabla$ is the covariant derivative in $g_{\alpha \beta}$ and $n_\alpha$ is the normal to $\Sigma$. The Levi-Civita connections of $g_{A B}$ and $\tilde{g}_{A B}$ are related by:
\be
\Gamma^B{}_{AC} = \tilde{\Gamma}^B{}_{AC}- \frac{2}{r}\left(\delta_{Cr}\, \delta^B_A +\delta_{Ar}\, \delta^B_C - g^{Br} g_{AC} \right)
\ee
For the calculation that follows, keep in mind the following: $e^\alpha_\phi=\delta^\alpha_\phi$ , $e^\alpha_\theta=\delta^\alpha_\theta$, $h^{\phi \phi} = g^{\phi \phi}$ , $h^{\theta \theta} = g^{\theta \theta}$, $h^{\lambda \lambda} = (g_{A B}e^A_\lambda e^B_\lambda)^{-1} $\,, $n_\alpha e^\alpha_a =0$. Also, notice that $n_\alpha$ and $e^\alpha_\lambda$ can be replaced by $n_A$ and $e^A_\lambda$ when contracted since they have vanishing angular components. 

We then have
\begin{eqnarray}
-K &=& -h^{ab} e^\alpha_a e^\beta_b \nabla_\alpha n_\beta  \nonumber \\ 
&=&   n_B (h^{ab} e^A_a \nabla_\alpha e^B_b) \nonumber \\
&=& n_B (g^{\phi \phi} \Gamma^B{}_{\phi \phi} + g^{\theta \theta} \Gamma^B{}_{\theta \theta} + h^{\lambda \lambda} e^A_\lambda \nabla_A \, e^B_\lambda  ) \nonumber \\ 
&=&  n_B (g^{\phi \phi} \Gamma^B{}_{\phi \phi} + g^{\theta \theta} \Gamma^B{}_{\theta \theta} \nonumber \\
&{}&\phantom{n_B (g^{\phi \phi} \Gamma^B{}_{\phi \phi} + g^{\theta \theta}}+ \frac{2}{r} h^{\lambda \lambda} g^{Br}g_{AC} e^A_\lambda e^C_\lambda  ) \nonumber \\
&=&  n_B g^{Br} (-g^{\phi \phi} \frac{g_{\phi \phi,r}}{2}   - g^{\theta \theta} \frac{g_{\theta \theta,r}}{2} + \frac{2}{r}) \nonumber \\
&=&  n_B g^{Br} (-\frac{1}{r}   - \frac{1}{r} + \frac{2}{r}) \nonumber \\
&=& 0 
\end{eqnarray}
where we used $\Gamma^B{}_{\phi \phi}=- \frac{1}{2} g^{Br}g_{\phi \phi,r}$, $\Gamma^B{}_{\theta \theta}=- \frac{1}{2} g^{Br}g_{\theta \theta,r}$ and that the surfaces are defined as $\Sigma \sim \gamma \times S^2$ with $\gamma$ a solution of eq.~\eqref{eq:geodEq}.

\bibliographystyle{JHEPs}
\bibliography{references}
\phantom{}

\end{document}